# Absolute cross sections for the dissociation of hydrogen cluster ions in high-energy collisions with helium atoms

S. Eden, J. Tabet, K. Samraoui [1], S. Louc, B. Farizon, M. Farizon, S. Ouaskit [1],

*Institut de Physique Nucléaire de Lyon, IN2P3-CNRS et Université Claude Bernard Lyon 1, 43, boulevard du 11 Novembre 1918, 69622 Villeurbanne Cedex, France*

and T. D. Märk

*Institut für Ionenphysik, Leopold Franzens Universität, Technikerstrasse 25, A-6020 Innsbruck, Austria*

[1] Permanent address: *Université Hassan II, Faculté des Sciences II, B.P. 6621-sidi Othemane, Casablanca, Morocco*

## Abstract

Absolute dissociation cross sections are reported for $H_n^+$ clusters of varied mass ($n$ = 3, 5, ..., 35) following collisions with He atoms at 60 keV / amu. Initial results have been published in a previous brief report for a smaller range of cluster sizes [Ouaskit *et al.*, *Phys. Rev. A* **49**, 1484 (1994)]. The present extended study includes further experimental results, reducing the statistical errors associated with the absolute cross sections. The previously suggested *quasi-linear* dependence of the $H_n^+$ dissociation cross sections upon $n$ is developed with reference to expected series of geometrical shells of $H_2$ molecules surrounding an $H_3^+$ core. Recent calculations identify $n$ = 9 as corresponding to the first closed $H_2$ shell [e.g. Štich *et al.*, *J. Chem. Phys.* **107**, 9482 (1997)]. Recurrence of the distinct characteristics observed in the dissociation cross section dependence upon cluster size around $n$ = 9 provides the basis for the presently proposed subsequent closed shells at $n$ = 15, 21, 27, and 33, in agreement with the calculations of Nagashima *et al.* [*J. Phys. Chem.* **96**, 4294 (1992)].

# 1. Introduction

Neutral and ionic hydrogen clusters are of major fundamental interest as they represent the simplest examples of molecular clustering. Furthermore, the mechanisms for $H_n^+$ cluster growth and fragmentation are relevant to a number of areas of applied physics. Cationic hydrogen clusters play a significant role in the stratosphere [1] where molecules readily nucleate about ions due to the long-range attractive forces. As a dominant ion in many hydrogen discharges and plasmas, $H_3^+$ is expected to be abundant in the interstellar medium [2]. The presence of $H_3^+$ ions provides the seed for $H_n^+$ cluster formation.

The structure of positively charged hydrogen clusters differs significantly from the neutral case. The latter constitutes a compact ensemble of $H_2$ molecules bound weakly by Van der Waals interactions [3], whereas positively charged hydrogen cluster ions are characterized by two additional effects: the delocalization of the electron hole and the polarization of $H_2$ molecules according to the London dispersion force. Thus, the *extra* proton leads to an increase in the stability of the system.

Ionic hydrogen clusters have been studied using a wide range of experimental and theoretical techniques [2-44]. Irrespective of the means of production, $H_n^+$ clusters comprising an uneven number of atoms have been observed to be by far the dominant species. Clampitt and Gowland [5] detected clusters produced by electron impact on a solid hydrogen target and proposed $H_3^+(H_2)_p$ as the generalized form, where $p$ is the number of $H_2$ molecules within the cluster. Following enthalpy measurements for the $H_n^+ \rightarrow H_{n-2}^+ + H_2$ reaction, Hiraoka [19] suggested the formation of successive $H_2$ shells for $p$ = 3, 6, 8, and 10 ($H_9^+$, $H_{15}^+$, $H_{19}^+$, and $H_{23}^+$) with closed shells corresponding to more stable configurations. Approximate binding energies were reported to range from 0.3 eV ($H_5^+$) to 0.02 eV ($H_{21}^+$) for the furthest $H_2$ molecule in the clusters [19]. Bae *et al.* [3] detected $H_3^+(H_2)_p$ clusters grown onto $H_3^+$ ions consistent with complete shells for $p$ = 3 and 6 ($H_9^+$ and $H_{15}^+$). The vibrational pre-dissociation spectra for hydrogen cluster ions up to $p$ = 5 ($H_{13}^+$) recorded around 4000 cm$^{-1}$ by Okumura *et al.* [16, 17] also support a closed $H_2$ shell for $p$ = 3 ($H_9^+$).

Relatively few experiments have been carried out concerning collision induced dissociation of hydrogen cluster ions. Van Lumig and Reuss [7-9] reported the dissociation of hydrogen cluster ions by low energy collisions (200-850 eV) with He, Ar, and $N_2$. For $N_2$ impact, $H_3^+(H_2)_p$ ($p \leq 9$) dissociation cross sections were observed to increase with cluster size according to a suggested 2/3 power law [7]. However, for He impact over a wider cluster size range ($p \leq 19$), a linear increase in dissociation cross section with cluster size was observed [8]. Indeed, for the smaller clusters ($p < 9$) a 2/3 power law dependence is only slightly different from a linear dependence. Van Lumig and Reuss [8] identified the $H_9^+$, $H_{15}^+$, and $H_{27}^+$ configurations as being particularly stable ($p = 3, 6,$ and $12$).

Chevallier *et al.* [18] studied the break-up of $H_n^+$ clusters in collisions with argon atoms. The dissociation cross section measurements were interpreted in terms of a dependence upon $n^{2/3}$, although the Ar impact dissociation cross sections for $H_{15}^+$ and $H_{19}^+$ are lower than expected. In the publication preceding the present work, Ouaskit *et al.* [27] suggested a *quasi-linear* increase with cluster size for the dissociation cross sections of $H_3^+(H_2)_p$ clusters ($p = 0$-14) following collisions with helium atoms. Ouaskit *et al.* [27] observed no evidence for shell effects.

Theoretical studies of hydrogen cluster ions have been carried out with increasingly regularity over recent years. Yamabe *et al.* [12] confirmed that the $H_3^+(H_2)_p$ ($p = 1$-4) clusters are dominated by the strongly bound regular $H_3^+$ triangle on the basis of self-consistent-field (SCF) calculations. Hirao and Yamabe [14] carried out SCF and configuration interaction (CI) calculations on the shell structure of $H_3^+(H_2)_p$ ($p = 1$-5) clusters and identified $p = 5$ ($H_{13}^+$) as being particularly stable. Nagashima *et al.* [25] determined *magic numbers* in the stability of $H_3^+(H_2)_p$ ions at $p = 3, 6, 9, 12,$ and $15$ ($H_9^+$, $H_{15}^+$, $H_{21}^+$, $H_{27}^+$, and $H_{33}^+$) using a classical Monte Carlo simulation and the annealing method. The *ab initio* CI calculations of Diekmann *et al.* [30], however, are consistent with full shells occurring at $p = 3, 5,$ and $9$ ($H_9^+$, $H_{13}^+$, and $H_{21}^+$).

By application of SCF and CI methods, Yamaguchi *et al.* [20] calculated the equilibrium geometry of the $H_3^+(H_2)$ cluster to have $C_{2v}$ symmetry with one of the atoms in the $H_3^+$ group bound to the $H_2$ molecule. The potential energy surface of the cluster has since been modeled by Prosmiti *et al.* [41] using perturbation theory. The geometrical structures and energetics of $H_7^+$ and $H_9^+$ have also been calculated by Yamaguchi *et al.* [15] and those of $H_3^+(H_2)_p$ ($p = 1$-6) by Farizon *et al.* [21-23, 35] using an *ab initio* Hartree-Fock (HF) technique. The distance between the protons within the $H_3^+$ core was reported to be 0.875 Å and that between the protons which form $H_2$ molecules to be of the order of 0.74 Å. The distance between the $H_2$ molecules was calculated as 1.6 Å for clusters of size less than or equal to $p = 3$. The larger clusters were described as a nucleation of $H_2$ molecules around a weakly deformed $H_9^+$ core. The distance between these additional molecules and the $H_3^+$ core was calculated to be of the order of 2.8 Å. These results are in agreement with the quantum Monte-Carlo calculations of Štich et al. [31, 32] for the cluster sizes $p = 0$-$3$ and 12. As expected, the structures derived by Štich et al. [32] correspond to a full $H_2$ shell at $p = 3$.

The density fluctuation calculations of Farizon et al. [35] suggest that any apparently high stability of the $H_{15}^+$ cluster cannot be attributed to a minimum energy configuration for a symmetric closed shell of 6 $H_2$ molecules around the $H_3^+$ core, as had been suggested by Hiraoka [19]. Accordingly, on the basis of ab initio SCF and HF calculations, Barbatti et al. [36] proposed $H_{15}^+$ to be neither a full shell nor any more stable than $H_{13}^+$ or $H_{17}^+$. However, the same authors [42] have since reported that $p = 6$ and 9 correspond to the closure of shells, in agreement with Nagashima et al. [25]. Similarly, the density functional theory (DFT) analysis of $H_3^+(H_2)_p$ ($p \leq 9$) carried out by Chermette and Ymmud [40] suggests $p = 6$ to be a *magic number* for cluster stability, although the effect may be subtle. Most recently, the structures, energetics, and anharmonic vibrational frequencies of small clusters ($p = 1$-$4$) were calculated using coupled-cluster perturbation and SCF methods [43, 44]. Prosmiti et al. [44] provided further evidence for a shell structure for $H_3^+(H_2)_p$ clusters, with $p = 3$ corresponding to the first closed shell.

The present work provides the most precise set of dissociation cross sections for fast $H_n^+$ impact with a neutral atomic target. Furthermore, data is reported for a wider range of cluster sizes than previously measured [18, 27]. The kinetic energies of the ionic cluster projectiles correspond to a velocity of $3.39 \times 10^6$ ms$^{-1}$, of the order of the Bohr velocity (c / 137 = $2.19 \times 10^6$ ms$^{-1}$). The relative velocity of the projectile and target atom thus approximates that of the electrons within the cluster. The resultant high efficiency for electronic excitation and ionization processes combined with the relatively weak binding of the clusters means that any collision around this energy has a high probability of initiating dissociation. Thus, considering also the short timeframe for collisions compared to nuclear motion, dissociation cross sections for collisions around the Bohr velocity can be expected to have a close relationship with the spatial distribution of the cluster constituents.

## 2. Experimental

The experimental technique is described in the previous publication [27]. Briefly, neutral clusters formed in a cryogenic source and ionized by electron impact are accelerated by means of a Cockcroft-Walton accelerator followed by a radio frequency quadrupole (RFQ) post-accelerator, as described by Gaillard *et al.* [45]. Following momentum analysis in a magnetic field, the angular dispersion of the cluster beam is limited to ± 0.8 mrad by two collimating apertures. The beam then passes between two parallel plates across which a deflecting voltage can be applied. Thus the beam can either be allowed to continue to the target or be directed to a surface-barrier detector in order to measure the incident cluster ion intensity. The intensity is limited to ~ 1000 cluster ions per second in order to avoid the overlap of signals either from this detector or from the undissociated cluster detector downstream.

The target is a gas jet formed by the expansion under vacuum of helium introduced into the collision chamber via a capillary tube. A detailed description of the target jet is given by Ouaskit *et al.* [46]. The capillary tube is mounted on a micrometric gonimeter which enables the jet to be translated in all directions (X, Y, Z) in steps of 50 μm. By means of a differential pumping system, the residual gas pressure close to the target jet and along the beamline is maintained at $7 \times 10^{-6}$ Torr.

The transmitted beam is analyzed in a magnetic field in a detection chamber located 1 m after the target, corresponding to a time of flight of 0.3 μs. The undissociated clusters are detected using a movable surface-barrier detector. It is worth noting that the signal from the surface-barrier detector is different for a molecular ion including oxygen, carbon, or nitrogen from that for a hydrogen cluster of the same mass and velocity [47].

In order to gain an absolute dissociation cross section from measurements of the fraction of $H_n^+$ clusters transmitted undissociated to the detection chamber, it is necessary to determine the thickness of the target jet. This depends upon the inlet pressure and the position of capillary with respect to the incident beam; parameters measured to a minimum precision of 0.3% and 4%, respectively. The transmitted fraction of $H_n^+$ clusters, F(*x*) is related to the thickness of the target, *x* by the equation below.

$$F(x) = F(0) \exp(-\sigma_d(H_n^+)\, x) \qquad \text{equation 1}$$

Where $F(0)$ is the transmitted fraction of $H_n^+$ clusters without the presence of any target gas and $\sigma_d(H_n^+)$ is the dissociation cross section. The absolute calibration technique is described in detail elsewhere [46].

The positioning of the capillary jet with respect to the direction of the incident beam is carried out once per assembly of the experiment and can introduce a systematic error for the corresponding set of data. The previous work [27] was carried out for a single preparation of the experimental system and the relative error of ± 8% was associated with target thickness uncertainty and fluctuations in the cluster beam intensity. The present results correspond to a number of separate preparations of the experimental system. Therefore, considering possible variation in the positioning of the jet, the maximum error on a single absolute cross section measurement, $\Delta\sigma_i$ is estimated to be ± 15%. However, the error is reduced statistically by averaging measurements repeated for different preparations of the system. Accordingly the errors on the present averaged cross sections, $\Delta\langle\sigma_d\rangle$ are calculated using the equation below.

$$\Delta\langle\sigma_d\rangle = \Delta\sigma_i / \sqrt{N} \qquad \text{equation 2}$$

Where N is the sample size. It is worth noting that the errors calculated using equation 2 are of the same order as the corresponding standard deviations (see table 1).

## 3. Results and Discussion

Dissociation cross sections have been measured following collisions with helium targets at 60 keV / amu for cluster ions comprised of between 5 and 35 atoms (odd numbers only) and for the molecular ion $H_3^+$. Figure 1 shows the complete set of results, including those presented in the previous short report [27]. The average dissociation cross sections are plotted in figure 2. In both figures, dissociation cross sections are plotted as a function of $p$, the number of $H_2$ molecules in the cluster. The corresponding values are listed in table 1. As expected, the dissociation cross sections increase significantly with cluster size: from $(2.77 \pm 0.11) \times 10^{-16}$ cm$^2$ for the molecular ion $H_3^+$ to $(44.98 \pm 6.75) \times 10^{-16}$ cm$^2$ for the $H_{35}^+$ cluster ($p = 16$).

The only available independent absolute cross sections which are comparable to the present data were measured by Wolff *et al.* [26] for $H_3^+$ dissociation in collisions with a helium target at velocities ranging from 160 to 1230 keV / amu. Wolff *et al.* [26] demonstrated an approximately linear dependence of the reciprocal of the $H_3^+$ dissociation cross section upon the square of the impact velocity in collisions with Ne, Ar, Xe, and He. This relation is associated with a simple analytical formula for electron loss in atomic systems developed by Meron and Johnson [48]. The reciprocal of the presently measured dissociation cross section of $H_3^+$, $(3.6 \pm 0.2) \times 10^{15}$ cm$^{-2}$, lies within the error boundaries of the extrapolation to (60 keV / amu)$^2$ of the weighted linear fit (Origin Pro™ Version 7) to the data of Wolff *et al.* [26], i.e., $(5.3 \pm 1.7) \times 10^{15}$ cm$^{-2}$. Thus the present absolute dissociation cross sections are consistent with the previous absolute measurements at higher energies [26].

Theoretical calculations for the structure of $H_n^+$ clusters [30] demonstrate that the electrons are localized on the $H_2$ molecules. Therefore it is interesting to consider the measured dissociation cross sections of $H_n^+$ clusters in the context of those for an $H_3^+$ ion or an $H_2$ molecule in isolation. As noted in the previous brief report [27], an initial basis for comparison is to sum the dissociation cross sections of the $H_3^+$ and $H_2$ elements in the system (equation 3).

$$\sigma_d(H_n^+) = p\,\sigma_d(H_2) + \sigma_d(H_3^+) \qquad \text{equation 3}$$

The measured cross section for the dissociation of the molecular ion, $\sigma_d(H_3^+)$, is given above. Whereas it was previously [27] necessary to approximate $\sigma_d(H_2)$ at 60 keV / amu using Bouliou *et al.*'s total ionization cross section $\sigma_I(H_2)$ [49], we can now use the recently measured $\sigma_d(H_2)$, $(3.15 \pm 0.47) \times 10^{-16}$ cm$^2$ [50]. Equation 3 is represented as a black dashed line in figure 2 with the error boundaries shown as dot-dashed lines.

The continuous line in figure 2 corresponds to a weighted linear fit for the experimental values. The gradient of the weighted fit, $(2.34 \pm 0.05) \times 10^{-16}$ cm$^2$ per $H_2$ molecule, is distinctly lower than the measured cross section for the dissociation of $H_2$ in isolation [50]. This seems at first surprising as, due to the weak inter-molecular binding energies, dissociation of the cluster may be initiated by an electronic excitation which is *non-dissociative* for the excited constituent molecule. However the low $\sigma_d(H_n^+)$ measurements compared to equation 3 can be rationalized in terms of screening; the cross sections associated with hydrogen molecules situated towards the *back* of the cluster with respect to the collision may be partially hidden by those towards the *front*.

Further insight into the evolution of $\sigma_d(H_n^+)$ with $n$ can be gained by consideration of vibrational excitation and, in particular, its dispersion throughout the cluster. For this purpose we can separate vibrational excitations according to their inter- or intra-molecular character. By increasing the time-averaged cluster volume, inter-molecular vibrations act to reduce the screening effect described above. This reduced screening, tending to increase cluster dissociation cross sections, may be expected to affect larger clusters to a greater extent due to their multiple degrees of freedom for inter-molecular vibration. However, as such vibrations can also cause dissociation in weakly-bound clusters, many possible $H_3^+(H_2)_p$ excited states may be too short-lived to exist at any significant concentration within the incident beam.

Intra-molecular vibration may play a role which tends to reduce the dissociation cross sections of $H_3^+(H_2)_p$ clusters. In particular, the $H_3^+$ core is understood to be quenched in cluster formation, leaving the ion in the ground state for intra-molecular (inter-nuclear) vibrations. Thus $H_3^+$ within a cluster is more dense than its isolated counterpart, and so is expected to have a lower dissociation cross section. From this perspective, it is interesting to note that the scattering effects observed for $H_3^+(H_2)_p$ collisions with an atomic target at 30-80 keV / amu are of the same order of magnitude as those which would be obtained for molecular beams of $H_3^+$ and $H_2$ in their fundamental state [24 and ref. therein].

Figure 3 shows the average dissociation cross sections as a function of the number of molecules in the cluster on a log-log scale. The line fitted to the experimental values for $p = 1 - 7$ has a gradient of 0.652 ± 0.014. This suggests a possible dependence of the cross section upon $p^{2/3}$, broadly consistent with the simple approximation of spherical $H_n^+$ clusters with uniform density proposed by Van Lumig and Reuss for $N_2$ [7] collisions. The volume of such a cluster would increase proportionally with the number of constituent molecules, and the radius and cross section according 1/3 and 2/3 power laws, respectively. The description is based upon the idea of an apparent dissociation cross section related to the spatial distribution of $H_2$ molecules. This distribution depends upon the geometric form averaged over all possible orientations with respect to the collision and upon the opacity of the cluster, which itself depends upon the structure and the collision process. The increase in $\log_{10} \sigma_d / \log_{10} p$ for larger $p$, clearly demonstrated in figure 3 for $p > 7$, may be understood in the context of Van Lumig and Reuss' rationale by the consideration of an additional effect due to $H_2$ molecules being more weakly bound further from the ionic core [32]. As the corresponding fall in overall cluster density with increasing $p$ will tend to reduce the overall screening effect, the measured dissociation cross section can be expected to increase at a faster rate with respect to $p$ than could be explained by a simple 2/3 power law.

The idea of variation in screening and the separation of $H_2$ molecules with $p$ suggests that shell effects may be visible in our dissociation cross section analysis. The recent calculations of Štich *et al.* [32] support the previous work indicating that the distance of $H_2$ molecules from the ionic core is distinct for each shell. Therefore we could expect a series of effective screening *regimes* corresponding to the occupation of successive shells with increasing $p$. Such regimes may be visible as distinct sub-regions of $\log_{10} \sigma_d$ dependence upon $\log_{10} p$.

The proposal shown in figure 4 for shell structure apparent in the present data is based upon the observation of distinct changes in the relation between $\log_{10} \sigma_d$ and $\log_{10} p$. To help identify such changes, Origin Pro™ Version 7 was used to evaluate the quality of weighted fits to $\log_{10} \sigma_d$ against $\log_{10} p$ with respect to the deviation of points in a series from a straight line. Importantly, this convenient system for the identification of cluster size steps which may involve a significant change in the arrangement of molecules in the cluster (i.e. the beginning of a new shell) distinguishes the $p = 3-4$ step as being significant; the linearity of $\log_{10} \sigma_d$ against $\log_{10} p$ in the range $p = 1-3$ ($H_5^+ - H_9^+$) is strikingly superior to that of any other possible series in the present data range. Accordingly, the first shell is widely accepted to close at $p = 3$ [3, 8, 16, 17, 19, 21-23, 25, 30, 32, 35, 44]. The subsequent domains of cluster size which we propose as shells on the basis of particularly good linearity for $\log_{10} \sigma_d$ against $\log_{10} p$ are $p = 4-6$ ($H_{11}^+ - H_{15}^+$ in agreement with references [8, 19, 25, 40, 42]), $p = 7-9$ ($H_{17}^+ - H_{21}^+$ [25, 42]), $p = 10-12$ ($H_{23}^+ - H_{27}^+$ [8, 25]), and possibly $p = 13-15$ ($H_{29}^+ - H_{33}^+$ [19, 25, 35]).

Figure 4 shows that the passage from $p = 3$ to $p = 4$ ($H_9^+$ to $H_{11}^+$) is characterized by the $p = 4$ point lying below the projected $p$ = 1-3 straight line. This seems at first surprising as cluster sizes corresponding to full shells would typically be expected to demonstrate increased stability through a tendency towards lower dissociation cross sections [7]. However, the relatively small increase in dissociation cross section for the step $p$ = 3-4 may be rationalized according to a hypothesis based upon vibrational excitation within closed shells. In the same way that the $H_3^+$ core of a cluster is understood to be in the ground state for intra-molecular vibration, the *intra-shell* (inter- and intra-molecular within the closed shell) vibrations of $H_9^+$ within $H_{11}^+$ may be restricted compared to $H_9^+$ in isolation. Such reduced vibration of $H_9^+$ would act to reduce the dissociation cross section of the $H_{11}^+$ cluster as a whole. This suggests a complimentary technique to look for shell breaks in the present data; a cluster size corresponding to the beginning of a shell may exhibit a dissociation cross section which is lower than would be expected by comparison with the previous cluster sizes. Accordingly, for each of the steps suggested above, figure 4 demonstrates that the values for $\log_{10} \sigma_d$ proposed to correspond to the beginning of a new shell lie below the projected straight line from the previous proposed shell. It should be noted that these discontinuities are not only apparent on a log-log scale; the same points are visibly lower than the projected relations suggested by the previous cluster sizes when $\sigma_d$ is plotted against $n$ or $p$ (see figure 2).

## 4. Conclusions

The most reliable and complete set of absolute cross sections are reported for cationic hydrogen cluster dissociation by high-energy atomic impact. The improved statistics reveal structure in the relation between dissociation cross section and cluster size which was not observable in the only previous absolute measurements [27].

The present analysis is based upon the widely accepted concept of a shell structure with the first full $H_3^+(H_2)_p$ shell occurring at $p$ = 3. However, while $p$ = 6 has been proposed by a number of previous authors, the *magic numbers* corresponding to closed $H_2$ shells above $p$ = 3 are a source of disagreement in the literature. The present interpretation of the data favors a series of shells comprising three $H_2$ molecules, as proposed by Nagashima *et al.* [25].

## Acknowledgements

The authors are grateful for the expert technical support of R. Genre, J. Martin, R. Filiol, J. P. Lopez, and H. Mathez. Partial financial support was provided by the FWF, Wien, Austria, the French, Austrian, and Morocco governments, the EU Commission (Brussels), through the Amadee and PICS 2290 programs, the CNRS-CNRST (n°17689) convention and the FP6 Marie Curie Fellowship (IEF RADAM-BIOCLUS). The *Institut de Physique Nucléaire de Lyon* is part of IN2P3-CNRS, the French national research institute for nuclear and particle Physics.

## Figure Captions

Figure 1: Dissociation cross sections for $H_3^+(H_2)_p$ clusters in 60 keV / amu collisions with helium atoms, $\sigma_d$, plotted as a function of $p$, the number of $H_2$ molecules per cluster in the range $0 \leq p \leq 16$ ($3 \leq n \leq 35$). The scatter diagram shows the complete set of results taken since the beginning of the experiment, including those presented in the previous short report [27].

Figure 2: Averaged dissociation cross sections for $H_3^+(H_2)_p$ clusters in 60 keV / amu collisions with helium atoms plotted as a function of $p$, the number of $H_2$ molecules in the clusters. The errors on the averaged cross sections are calculated using $\Delta<\sigma_d> = \Delta\sigma_i/\sqrt{N}$, where $\sigma_i$ is $\pm$ 15% and N is the sample size. Black dashed line: $\sigma_d = p\,\sigma_d(H_2) + \sigma_d(H_3)$ using $\sigma_d(H_2) = (3.15 \pm 0.47) \times 10^{-16}$ cm$^2$. Gray dashed lines: error boundaries for $\sigma_d = p\,\sigma_d(H_2) + \sigma_d(H_3)$. Continuous line: weighted linear fit to the experimental values.

Figure 3: Log-log plot of the averaged dissociation cross sections for $H_3^+(H_2)_p$ clusters in 60 keV / amu collisions with helium atoms plotted as a function of $p$, the number of $H_2$ molecules. Dotted line: fitted to the experimental values for $p = 1$-7 (gradient $0.652 \pm 0.014$).

Figure 4: Log-log plot of the averaged dissociation cross sections for $H_3^+(H_2)_p$ clusters in 60 keV / amu collisions with helium atoms plotted as a function of $p$, the number of $H_2$ molecules. Lines: weighted linear fits for (from left to right) $p = 1$–3, $p = 4$-6, $p = 7$-9, $p = 10$-12, and $p = 13$-15.

## Table Captions

Table 1: Average dissociation cross sections for $H_n^+$ clusters following helium impact at 60 keV / amu.

Figure 1: Dissociation cross sections for $H_3^+(H_2)_p$ clusters in 60 keV / amu collisions with helium atoms, $\sigma_d$, plotted as a function of $p$, the number of $H_2$ molecules per cluster in the range $0 \leq p \leq 16$ ($3 \leq n \leq 35$). The scatter diagram shows the complete set of results taken since the beginning of the experiment, including those presented in the previous short report [27].

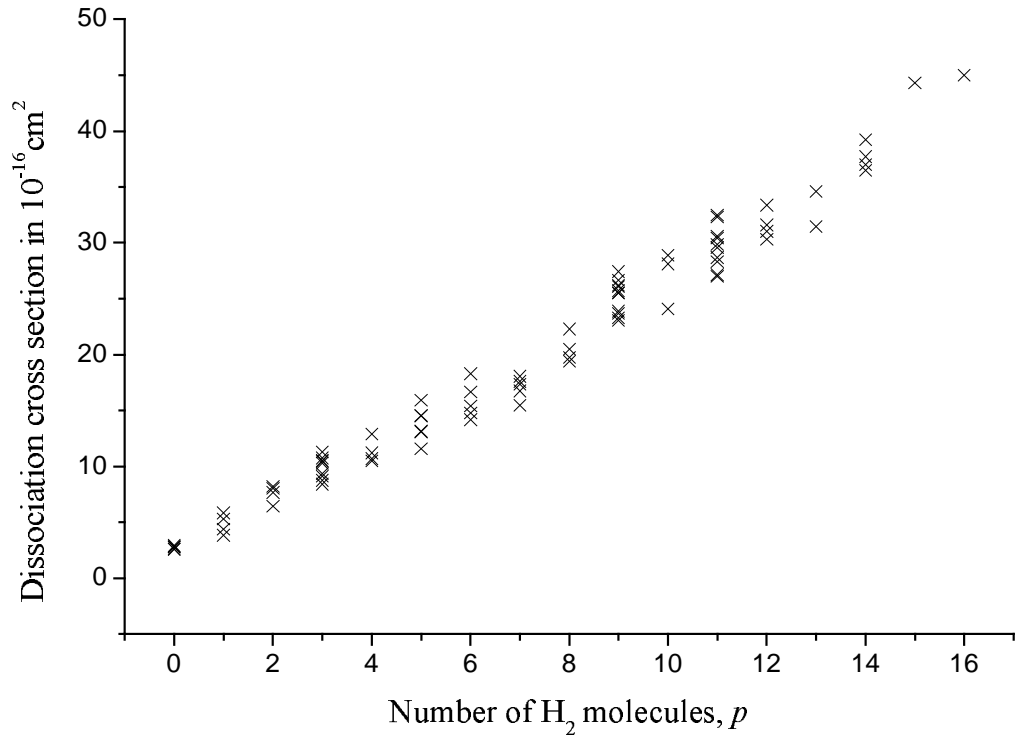

Figure 2: Averaged dissociation cross sections for $H_3^+(H_2)_p$ clusters in 60 keV / amu collisions with helium atoms plotted as a function of $p$, the number of $H_2$ molecules in the clusters. The errors on the averaged cross sections are calculated using $\Delta\langle\sigma_d\rangle = \Delta\sigma_i/\sqrt{N}$, where $\sigma_i$ is ± 15% and N is the sample size. Black dashed line: $\sigma_d = p\ \sigma_d(H_2) + \sigma_d(H_3)$ using $\sigma_d(H_2) = (3.15 \pm 0.47) \times 10^{-16}$ cm$^2$. Gray dashed lines: error boundaries for $\sigma_d = p\ \sigma_d(H_2) + \sigma_d(H_3)$. Continuous line: weighted linear fit to the experimental values.

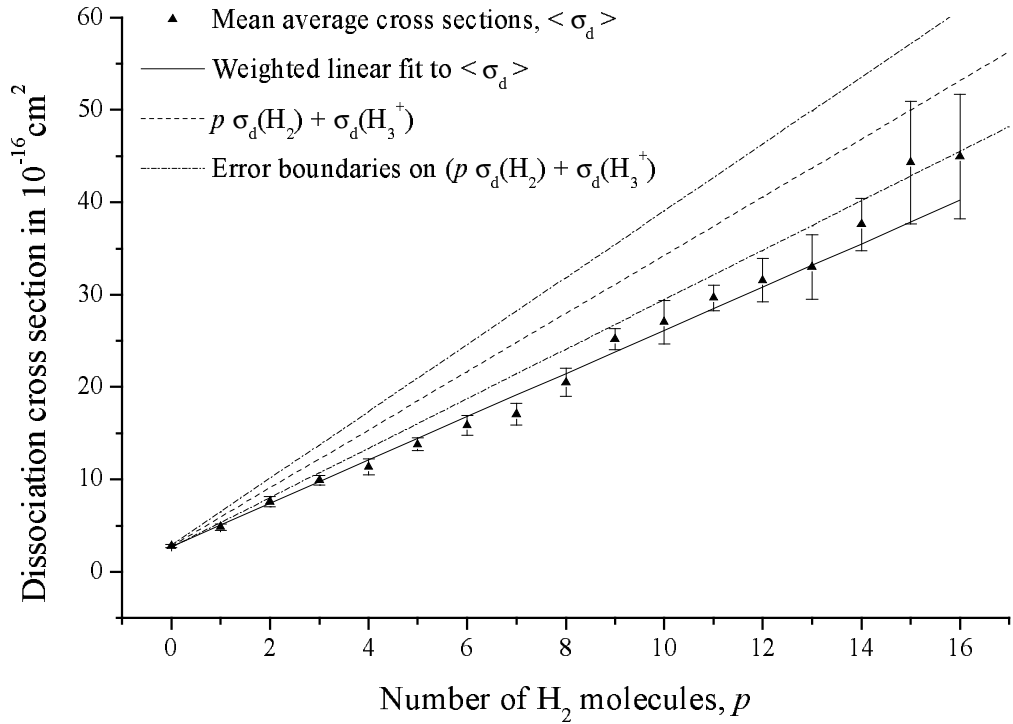

Figure 3: Log-log plot of the averaged dissociation cross sections for $H_3^+(H_2)_p$ clusters in 60 keV / amu collisions with helium atoms plotted as a function of $p$, the number of $H_2$ molecules. Dotted line: fitted to the experimental values for $p$ = 1-7 (gradient 0.652 ± 0.014).

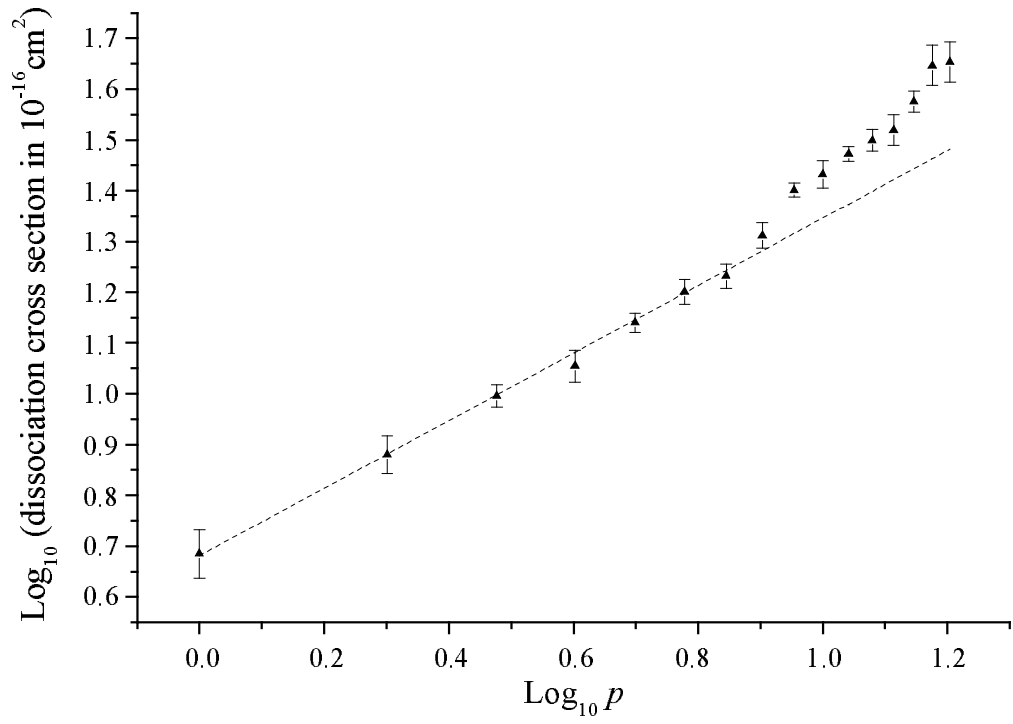

Figure 4: Log-log plot of the averaged dissociation cross sections for $H_3^+(H_2)_p$ clusters in 60 keV / amu collisions with helium atoms plotted as a function of $p$, the number of $H_2$ molecules. Lines: weighted linear fits for (from left to right) $p = 1-3$, $p = 4-6$, $p = 7-9$, $p = 10-12$, and $p = 13-15$.

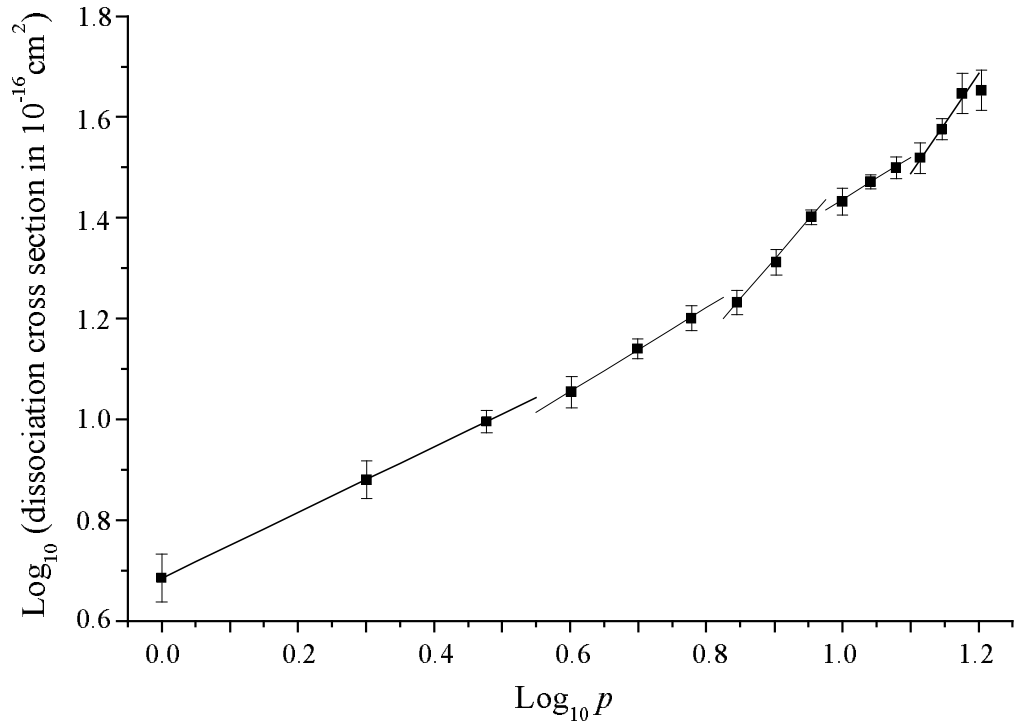

Table 1: Averaged dissociation cross sections for $H_3^+(H_2)_p$ clusters following helium impact at 60 keV / amu

| Cluster size, $n$ | Number of $H_2$ molecules, $p$ | Sample size, N | Average $\sigma_d$ in $10^{-16}$ cm$^2$ | $\Delta <\sigma_d>$ = $\Delta\sigma_i$ */ $\sqrt{N}$ | Standard deviation |
|---|---|---|---|---|---|
| 3 | 0 | 6 | 2.77 | ± 0.17 | 0.11 |
| 5 | 1 | 4 | 4.84 | ± 0.36 | 0.77 |
| 7 | 2 | 4 | 7.59 | ± 0.57 | 0.68 |
| 9 | 3 | 9 | 9.90 | ± 0.50 | 0.94 |
| 11 | 4 | 4 | 11.34 | ± 0.85 | 0.97 |
| 13 | 5 | 9 | 13.81 | ± 0.69 | 1.36 |
| 15 | 6 | 5 | 15.87 | ± 1.06 | 1.47 |
| 17 | 7 | 5 | 17.05 | ± 1.14 | 0.89 |
| 19 | 8 | 4 | 20.50 | ± 1.54 | 1.11 |
| 21 | 9 | 11 | 25.19 | ± 1.14 | 1.40 |
| 23 | 10 | 3 | 27.04 | ± 2.34 | 2.10 |
| 25 | 11 | 10 | 29.64 | ± 1.41 | 1.81 |
| 27 | 12 | 4 | 31.57 | ± 2.37 | 1.13 |
| 29 | 13 | 2 | 33.02 | ± 3.50 | 1.58 |
| 31 | 14 | 4 | 37.62 | ± 2.82 | 1.02 |
| 33 | 15 | 1 | 44.31 | ± 6.65 | - |
| 35 | 16 | 1 | 44.98 | ± 6.75 | - |
| * $\Delta\sigma_i$ is estimated at ±15% | | | | | |